\documentclass[sigconf,screen]{acmart}

\usepackage{algorithmic}
\usepackage{graphicx}
\usepackage{textcomp}
\usepackage{tcolorbox}
\usepackage{pifont}
\usepackage{xcolor}
\usepackage{soul}
\usepackage{booktabs}
\usepackage{multirow}
\usepackage{float}
\usepackage{url}
\usepackage{graphicx}
\usepackage{balance}
\usepackage{microtype}
\usepackage[numbered]{bookmark}
\usepackage{tcolorbox}
\usepackage{titlesec}
\usepackage{enumitem}
\usepackage{seqsplit}

\setlist[itemize,enumerate]{noitemsep, topsep=0pt, leftmargin=1.0em}
\usepackage{draftwatermark}
\SetWatermarkText{PREPRINT}
\SetWatermarkLightness{0.88}
\SetWatermarkScale{0.33}

\renewcommand\footnotetextcopyrightpermission[1]{}
\usepackage{listings}

\definecolor{javared}{rgb}{0.6,0,0} % for strings
\definecolor{javagreen}{rgb}{0.25,0.5,0.35} % comments
\definecolor{javapurple}{rgb}{0.5,0,0.35} % keywords
\definecolor{javadocblue}{rgb}{0.25,0.35,0.75} % javadoc
 
\definecolor{light-red}{rgb}{1,0.92,0.91}
\definecolor{light-green}{rgb}{0.9,1,0.93}

\acmConference[MSR 2022]{MSR '22: Proceedings of the 19th International Conference on Mining Software Repositories}{May 23–24, 2022}{Pittsburgh, PA, USA}

\newcommand{\RQA}{\textbf{RQ1}: To what extent do developers refactor code when removing technical debt?}

\newcommand{\RQB}{\textbf{RQ2}: What motivates a developer to refactor their source code to remove technical debt?}

\begin{document}


% \keywords{datasets, neural networks, gaze detection, text tagging}
\newcommand\Mark[1]{\textsuperscript#1}
\title{Refactoring Debt: Myth or Reality? An Exploratory Study on the Relationship Between Technical Debt and Refactoring}
% \title{On the Removal of Technical Debt Through Refactoring: An Exploratory Study}

% \author{\large Anthony Peruma\Mark{1}, Eman Abdullah AlOmar\Mark{1}, Christian D. Newman\Mark{1}, Mohamed Wiem Mkaouer\Mark{1}, Ali Ouni\Mark{2}}
% \affiliation{
%     \institution{\Mark{1}Rochester Institute of Technology, Rochester, NY, USA}
%     \institution{\Mark{2}ETS Montreal, University of Quebec, Montreal, QC, Canada}
% }
% \email {axp6201@rit.edu, eman.alomar@mail.rit.edu, cnewman@se.rit.edu, mwmvse@rit.edu, ali.ouni@etsmtl.ca} 

\author{Anthony Peruma}
\email{axp6201@rit.edu}
\affiliation{%
  \institution{Rochester Institute of Technology}
  \city{Rochester}
  \state{New York}
  \country{USA}
}

\author{Eman Abdullah AlOmar}
\email{eman.alomar@mail.rit.edu} 
\affiliation{%
 \institution{Rochester Institute of Technology}
 \city{Rochester}
 \state{New York}
 \country{USA}}

\author{Christian D. Newman}
\email{cnewman@se.rit.edu}
\affiliation{%
  \institution{Rochester Institute of Technology}
  \city{Rochester}
  \state{New York}
  \country{USA}
}

\author{Mohamed Wiem Mkaouer}
\email{mwmvse@rit.edu} 
\affiliation{%
 \institution{Rochester Institute of Technology}
 \city{Rochester}
 \state{New York}
 \country{USA}}
 
\author{Ali Ouni}
\email{ali.ouni@etsmtl.ca} 
\affiliation{%
  \institution{ETS Montreal, University of Quebec}
  \city{Montreal}
  \state{Quebec}
  \country{Canada}}

\renewcommand{\shortauthors}{Peruma et al.}

\begin{abstract}
To meet project timelines or budget constraints, developers intentionally deviate from writing optimal code to feasible code in what is known as incurring \textit{Technical Debt} (TD). Furthermore, as part of planning their correction, developers document these deficiencies as comments in the code (i.e., self-admitted technical debt or SATD). As a means of improving source code quality, developers often apply a series of refactoring operations to their codebase. In this study, we explore developers repaying this debt through refactoring operations by examining occurrences of SATD removal in the code of 76 open-source Java systems. Our findings show that TD payment usually occurs with refactoring activities and developers refactor their code to remove TD for specific reasons. We envision our findings supporting vendors in providing tools to better support developers in the automatic repayment of technical debt.
\end{abstract}

\maketitle

\section{Introduction}
\label{Section:introduction}

In 1992, Ward Cunningham coined the debt metaphor when explaining to product stakeholders the need to keep improving the quality of their software through refactoring \cite{cunningham1992wycash}. Since then, \textit{Technical Debt} (TD) has become a reference to any non-optimal code shipped to production with the promise of enhancing it in the next cycle, i.e., developers \textit{owe} users better software in a future release. In this context, \textit{Self-Admitted Technical Debt} (SATD) refers to developers deliberately admitting the existence of TD in their system by keeping track of it through inline documentation, which typically manifests in the form of source code comments that describe an anomaly of a code element (e.g., class, method, etc.) \cite{potdar2014exploratory,lenarduzzi2019towards}.

Since SATD is a declaration of the need to update the system, several studies have analyzed its removal to understand better how developers address the issues mentioned in the comments, and consequently manage and reduce TD \cite{Iammarino2019SATD,iammarino2021empirical,tan2022does,oizumi2020recommending,lenarduzzi2021systematic}. Yet, little is known about the extent to which refactoring contributes to the removal of the issues documented in the SATD comments. For example, in Listing \ref{Listing:intro}, the comment mentions the need to rename a method name as per its current implementation. Therefore, the fix of this SATD was performed through the application of the \textit{Rename Method} refactoring, where the method \texttt{getVertexName()} was renamed to \texttt{getName()}, with the commit message ``Replace confusing names on Vertex API''.

\vspace{2mm}
\begin{minipage}{\linewidth}
\begin{lstlisting}[caption=Code diff showing the renaming of a method due to its associated technical debt comment and the removal of the same comment \cite{intro_example}., label=Listing:intro, firstnumber = last, escapeinside={(*@}{@*)},escapechar=!]
  !\colorbox{light-red}{- public String getVertexName() }!
  !\colorbox{light-red}{- // FIXME rename to getName()}!
  !\colorbox{light-green}{+   public String getName() }!
\end{lstlisting}
\end{minipage}

\vspace{-4mm}
\subsection{Goal \& Research Questions}
The goal of this paper is to explore the co-occurrence of refactorings with the removal of SATD. We also distill the types of debt being addressed by extracting topics from the text of the comments. To this end, we answer the following Research Questions (RQs):
\begin{itemize}
    \item \textbf{\RQA} This RQ gives us insight into the volume of occurrence of technical debt in our dataset and examines the types of refactorings that developers frequently apply.
    
    \item \textbf{\RQB} Since SATD are indicators of shortcomings in the code, this RQ examines the repayment categories for which developers refactor their code.
\end{itemize}

% \ali{briefly describe with numbers the experiments (number of projects, number of studied SATD instances, etc.)}

% This paper provides promising preliminary findings that empirically and statistically show a strong association between refactoring and the repayment of TD through the removal of SATD comments. We also show various rationales for refactoring for debt payment. Finally, we provide the community with a structured dataset to further investigate the cause-effect relationship between refactoring and removed SATD comments. 

\section{Experiment Design}
\label{Section:experiment_design}
\begin{figure}[h]
 	\centering
 	\includegraphics[trim=0cm 0cm 0cm 0cm, width=1\linewidth]{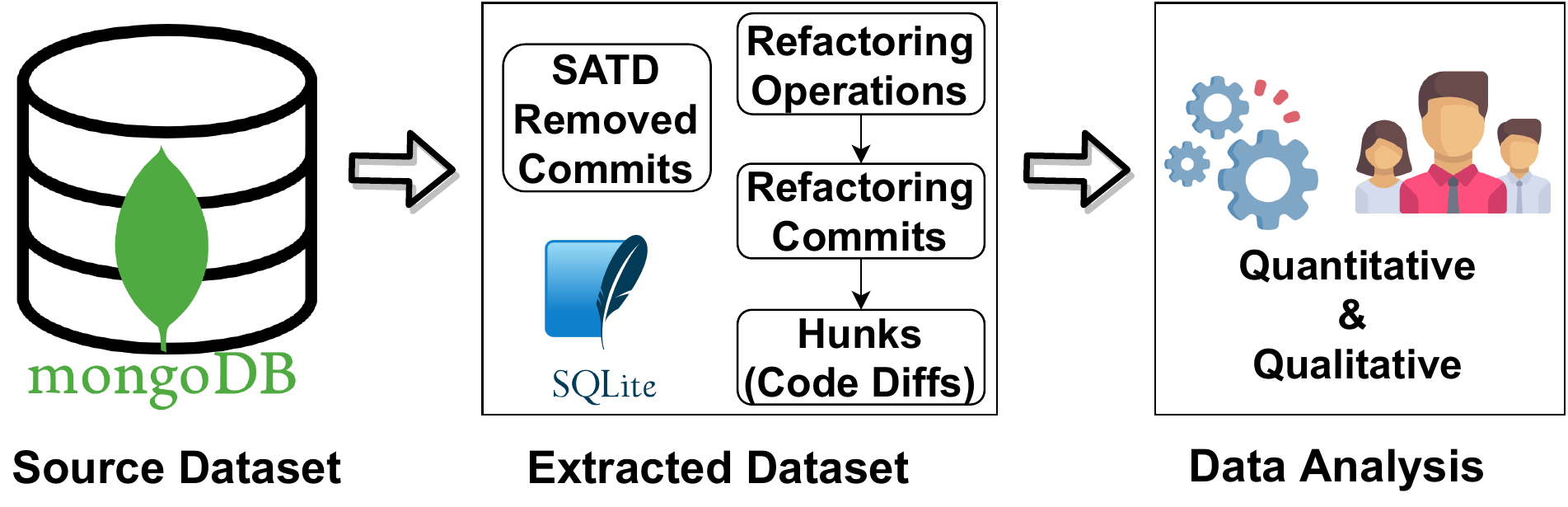}
 	\caption{Overview of our experiment design.}\vspace{-2.5mm}
 	\label{Figure:diagram_experiment}
\end{figure}

Figure \ref{Figure:diagram_experiment} outlines the experiment for our study. In the following subsections, we describe the elements and activities that were part of our methodology. Our replication package is available at \cite{ProjectWebSite}.

\vspace{-1.5mm}

\vspace{-2mm}
\subsection{Source Dataset}
\label{Section:experiment_design_source}
In this study, we utilize the \textit{SmartSHARK MongoDB Release 2.1 (full version)} dataset  \cite{Trautsch2021MSRSmartShark}. The dataset contains the commit details of 77 open-source Java projects that are part of the Apache ecosystem and utilize GitHub as their project repository. Furthermore, the dataset also contains the changed hunks (including diffs) of each file and utilizes RefDiff \cite{Silva2017Refdiff} and RefactoringMiner \cite{Tsantalis2018RMiner} to mine refactoring operations. Furthermore, the authors of the dataset indicate if a commit includes the removal of SATD by analyzing the code comment diffs in the source files associated with the commit. The authors examine the comment diffs for the presence or removal of the terms: ``TODO'', ``FIXME'', and ``XXX''.  

\subsection{Extracted Dataset}
Since the source dataset contains an abundance of data, some of which is not related to our study, we built custom scripts to extract data pertinent to our study (i.e., commits, hunks, refactorings) from the source dataset into an SQLite database for analysis. First, we extract all commits with a label showing SATD removal. Next, we extract all refactoring operations. However, due to the use of two refactoring mining tools, there are duplicate operations in the source data. Hence, our next step removes all duplicates by comparing the refactoring descriptions. After that, we select all commits associated with a refactoring operation. Finally, we extract all hunks (i.e., code diffs) in files associated with the extracted refactoring commits. Table \ref{Table:Dataset} is a summary of the extracted data. 

\begin{table}
\centering
\caption{Summary of the extracted data.}
\vspace{-3mm}
\label{Table:Dataset}
\begin{tabular}{@{}lr@{}}
\toprule
\multicolumn{1}{c}{\textbf{Item}}    & \multicolumn{1}{c}{\textbf{Value}} \\ \midrule
Count of total projects                    & 77                                 \\
Count of commits in all projects                    & 366,322                                 \\
Count of projects with SATD removed                    & 76                                 \\
Count of commits with SATD removed & 13,259                         \\ 
Count of refactoring operations      & 703,260                            \\
Count of commits with   refactorings & 67,582                             \\
Count of refactoring commits with SATD removed & 7,341                              \\ \bottomrule
\end{tabular}
\vspace{-4mm}
\end{table}

\subsection{Data Analysis}

Our analysis of the extracted data follows a mixed-methods approach, where we collect and analyze both quantitative and qualitative data \cite{Tashakkori1998Mixed}. This approach presents us with the opportunity to provide representative samples from the dataset to complement our findings. Our quantitative approach utilizes well-established statistical measures and custom code/scripts on our dataset to report trends and patterns. In Section \ref{Section:experiment_results}, we elaborate in detail on our analysis approach to answering each research question.

\section{Experiment Results}
\label{Section:experiment_results}

In this section, we report on the findings of our experiments by answering our RQs. To answer our RQs, \textbf{we utilize SATD as a proxy for technical debt}, and hence, the removal of such comments indicates that developers removed technical debt code from the same source code file. For each RQ, we first explain the primary motivation(s) and the approach we undertake to produce the results, then we present our findings.

\subsection*{\RQA}
\textbf{RQ1} provides a deep-dive quantitative-based examination of refactoring operations co-occurring with technical debt removal. To this extent, this RQ looks into the volume of occurrence of technical debt in our dataset and examines the types of refactorings that developers frequently apply.

% \begin{figure*}
%  	\centering
%  	\includegraphics[trim=2.7cm 6.8cm 0.5cm 3.7cm,clip,scale=0.8]{figures/CorrPlot_RefactoringOperations.pdf}
%  	\caption{Correlation matrix showing the relationship of applied refactoring operations in files with and without technical debt removal. Circles in red indicate a negative association, while blue is positive.} 
%  	\label{Figure:chart}
% \end{figure*}

% \subsubsection*{\RQAA}\ali{like we are assuming that developers do apply refactoring to remove TD. I think we are not sure about this so far.}
% This RQ quantitatively examines the occurrences of refactoring operations with technical debt removal. 
\begin{table}
\centering
\caption{Statistical summary of technical debt removal related to files in a commit and applied refactoring operations.}
\vspace{-2.5mm}
\label{Table:rq1_stats}
\begin{tabular}{crrrrr}
\hline
\textbf{Min.} & \multicolumn{1}{c}{\textbf{1st Qu.}} & \multicolumn{1}{c}{\textbf{Median}} & \multicolumn{1}{c}{\textbf{Mean}} & \multicolumn{1}{c}{\textbf{3rd Qu.}} & \multicolumn{1}{c}{\textbf{Max.}} \\ \hline
\multicolumn{6}{c}{\textit{Files in a refactoring commit having technical debt removed}} \\
\multicolumn{1}{r}{1} & 5 & 12 & 54.56 & 25 & 7593 \\ \hline
\multicolumn{6}{c}{\textit{Files in a refactoring commit without technical debt removed}} \\
\multicolumn{1}{r}{1} & 2 & 4 & 10.04 & 10 & 2315 \\ \hline
\multicolumn{6}{c}{\textit{Refactoring operations in a file with technical debt removal}} \\
\multicolumn{1}{r}{1} & 1 & 2 & 3.278 & 4 & 96 \\ \hline
\multicolumn{6}{c}{\textit{Refactoring operations in a file without technical debt removal}} \\
\multicolumn{1}{r}{1} & 1 & 1 & 2.774 & 3 & 161 \\ \hline
\end{tabular}
\vspace{-.4cm}
\end{table}

\begin{table*}
	\fontsize{7.5}{9}\selectfont
\centering
\caption{Distribution of top 10 refactoring operations on commits containing a single file with and without SATD removal.}
\vspace{-2.5mm}
\label{Table:rq1_operations}
%\resizebox{\textwidth}{!}{%
\begin{tabular}{lrrc|r|lrrc}
\hline
\multicolumn{4}{c|}{\textbf{Commit with one file with SATD removal}}                                                                                                                                                                                                                                                                         &  & \multicolumn{4}{c}{\textbf{Commit with one file without SATD removal}}                                                                                                                                                                                                                                                                      \\ \hline
\multicolumn{1}{c}{\textbf{\begin{tabular}[c]{@{}c@{}}Refactoring \\ Operation Type\end{tabular}}} & \multicolumn{1}{c}{\textbf{\begin{tabular}[c]{@{}c@{}}Total \\ Count\end{tabular}}} & \multicolumn{1}{c}{\textbf{Percentage}} & \multicolumn{1}{c|}{\textbf{\begin{tabular}[c]{@{}c@{}}Avg. Operation \\ Counts Per File\end{tabular}}} &  & \multicolumn{1}{c}{\textbf{\begin{tabular}[c]{@{}c@{}}Refactoring \\ Operation Type\end{tabular}}} & \multicolumn{1}{c}{\textbf{\begin{tabular}[c]{@{}c@{}}Total \\ Count\end{tabular}}} & \multicolumn{1}{c}{\textbf{Percentage}} & \multicolumn{1}{c}{\textbf{\begin{tabular}[c]{@{}c@{}}Avg. Operation \\ Counts Per File\end{tabular}}} \\ \hline
Extract Method                                                                                     & 289                                                                                 & 22.32\%                                 & 2.16                                                                                                    &  & Change Variable Type                                                                               & 6,519                                                                               & 21.67\%                                 & 2.39                                                                                                   \\
Change Variable Type                                                                               & 234                                                                                 & 18.07\%                                 & 2.27                                                                                                    &  & Extract Method                                                                                     & 6,188                                                                               & 20.57\%                                 & 2.36                                                                                                   \\
Rename Variable                                                                                    & 141                                                                                 & 10.89\%                                 & 1.60                                                                                                    &  & Rename Variable                                                                                    & 2,648                                                                               & 8.80\%                                  & 1.54                                                                                                   \\
Rename Method                                                                                      & 126                                                                                 & 9.73\%                                  & 1.66                                                                                                    &  & Rename Method                                                                                      & 2,473                                                                               & 8.22\%                                  & 1.78                                                                                                   \\
Rename Parameter                                                                                   & 99                                                                                  & 7.64\%                                  & 1.90                                                                                                    &  & Extract Attribute                                                                                  & 2,459                                                                               & 8.17\%                                  & 3.01                                                                                                   \\
Extract Attribute                                                                                  & 96                                                                                  & 7.41\%                                  & 5.05                                                                                                    &  & Extract Variable                                                                                   & 2,336                                                                               & 7.76\%                                  & 1.32                                                                                                   \\
Inline Method                                                                                      & 58                                                                                  & 4.48\%                                  & 2.07                                                                                                    &  & Change Return Type                                                                                 & 2,054                                                                               & 6.83\%                                  & 2.22                                                                                                   \\
Extract Variable                                                                                   & 54                                                                                  & 4.17\%                                  & 1.17                                                                                                    &  & Rename Parameter                                                                                   & 1,672                                                                               & 5.56\%                                  & 1.81                                                                                                   \\
Rename Attribute                                                                                   & 51                                                                                  & 3.94\%                                  & 1.59                                                                                                    &  & Rename Attribute                                                                                   & 896                                                                                 & 2.98\%                                  & 1.43                                                                                                   \\
Change Return Type                                                                                 & 43                                                                                  & 3.32\%                                  & 1.48                                                                                                    &  & Inline Method                                                                                      & 591                                                                                 & 1.96\%                                  & 1.70                                                                                                   \\
\textit{Other Operations}                                                                          & 104                                                                                 & 8.03\%                                  & \textit{N/A}                                                                                            &  & \textit{Other Operations}                                                                          & 2,248                                                                               & 7.47\%                                  & \textit{N/A}                                                                                           \\ \hline
\end{tabular}%
%}
\vspace{-3 mm}
\end{table*}

\noindent\textbf{Approach:} A limitation of the source dataset is that it only provides the commit associated with a refactoring operation and not the actual source file to which the refactoring is applied. Therefore, our examination is limited to the commit level. First, we look at all occurrences of refactoring commits having technical debt removed, and then (to mitigate false positives) we narrow our analysis to only refactoring commits containing a single file with technical debt removed. We opt for this decision as it allows us to analyze cases where the removed TD and the refactoring are co-located in the same code. When analyzing commits with one file, we select two groups of refactored files-- those that have technical debt removed and those that do not. %From these two groups, we examine the distribution of refactoring operations. 

% \vspace{1.3mm}
\noindent\textbf{Analysis 1: Volume of refactoring and technical debt removal.}
\newline
First, we observe that 76 of the 77 projects in our dataset contain technical debt removal and the application of refactoring operations. We encounter 7,341 commits having refactorings and technical debt removal. This is around 55.37\% of all commits labeled as having technical debt removed in the source dataset (with and without refactoring). As our study is on the relationship between technical debt removal and refactoring, we focus on the 7,341 refactoring commit instances. In contrast, 60,241 (or 89.14\%) refactoring commits do not show the removal of technical debt. Our dataset contains 395 (or 5.38\%) refactoring commits with only one file that exhibits the removal of technical debt. Conversely, there are 10,845 (or 18\%) refactoring commits with only one file that does not have technical debt removed. Our analysis also includes measuring the association between debt repayment and refactoring by performing an odds ratio (OR) test for each of the 76 systems. All 76 systems have an OR $>$ 1, with the value of 74 systems being statistically significant (i.e., $p$-value $<$ 0.05), showing \textit{a greater likelihood of debt repayment through refactoring activities}.

% \begin{tcolorbox}
% Across 74 projects, there is a statistically significant likelihood that a refactoring has been performed in the file experiencing the removal of a TD.
% \end{tcolorbox}

Next, we look at the volume files and refactoring operations. Since a commit can contain more than one file, an examination of commits shows the median number of files with technical debt removal in a refactoring commit is 12. In contrast, the median value in a refactoring commit without technical debt removal is 4. The complete statistical summary is in Table \ref{Table:rq1_stats}. Finally, we compare the differences between the number of refactoring operations in each file type (i.e., with and without technical debt removal). We utilize the non-parametric Mann Whitney U test as the data is not normally distributed. In this test, our null hypothesis (H\textsubscript{0}) is that the distribution of refactoring operations for the two groups is equal. The results of the test yield $p$-value $<$ 0.05, thereby rejecting H\textsubscript{0} and shows that a difference between the population is statistically significant.   

% These findings show that \textbf{\textit{developers refactor code when removing technical debt in isolation than when performing general code refactoring}}.   \ali{one question that come to my mind: how many refactoring commits does a TD require to be fixed?} % Furthermore, we observe that \textbf{\textit{all commits with technical debt removal also exhibit an application of one or more refactoring operations}}.

% \subsubsection*{\RQAB}
% This RQ aims to determine the common refactoring operations developer apply when removing technical debt and how they differ from non-technical debt removal updates.

% \noindent\textbf{Approach:} As previously mentioned, the source dataset is limited to only providing the commit associated with a refactoring operation and not the actual source file to which the refactoring is applied. Hence, since a commit can contain one or more files, \textit{our analysis is limited to commits with only one file to mitigate false positives}. As such, we select two groups of refactored files-- those that have SATD removed and those that do not. From these two groups, we examine the distribution of refactoring operations. 
% \vspace{2.0mm}
\noindent\textbf{Analysis 2: Refactoring operations typically co-occuring with technical debt removal.}
\newline
Shown in Table \ref{Table:rq1_operations} are the frequently occurring refactoring operations developers apply to a file when removing SATD and also the operations that are frequent in source files where there is no removal of SATD. Comparing these two categories shows that the top ten refactoring operations are the same, even though the percentage frequency of occurrence for some differ. Furthermore, we observe that the files without technical debt removal undergo more types of refactoring operations; 32 types against 21. Looking at the average number of operations in a file, we observe developers frequently apply the \textit{Extract Attribute} operation to the class followed by \textit{Change Variable Type} and \textit{Extract Method}. Next, we apply a Fisher's Exact Test to examine a statistically significant association between the average number of refactoring operations, of each refactoring type, applied to a file with and without technical debt. We utilize this test since we are comparing two categorical groups. In this test, variable independence is our null hypothesis (H\textsubscript{0}). The results yield a $p$-value $>$ 0.05, resulting in the acceptance of H\textsubscript{0}. Therefore, showing the developers apply the different refactoring operations to improve the quality of the code regardless if the developer is addressing technical debt concerns in the code.

\vspace{-1.5mm}
\begin{tcolorbox}[top=0.5pt,bottom=0.5pt,left=1pt,right=1pt]
\textbf{Summary for RQ1.}
The repayment of technical debt is an activity that developers often perform when refactoring code by applying a variety of refactoring operations, with \textit{Extract Attribute} frequently applied multiple times in a file. Furthermore, there is a significant difference in the volume of refactorings developers apply when removing technical debt compared to general maintenance activities. 
\end{tcolorbox}

\subsection*{\RQB}
The previous RQ shows that removing technical debt is often associated with a refactoring activity. However, we lack context around the rationale. As SATD-related comments indicate shortcomings in the existing code, removing such comments is usually an indicator that developers correct the deficiency. Hence, an analysis of these comments provides insight into the issues the developer is correcting by performing one or more refactoring operations. To this end, this RQ constructs a grouping of {\textit{technical debt categories developers resolve through refactoring}}. %In the first sub-RQ, we examine the set of frequently occurring terms in SATD comments.

%\subsubsection*{\RQBA}
%In this sub-RQ, we follow an approach similar to \citet{Peruma2021EMSEStackoverflow} to determine the typical terms as bigrams in the SATD comment to understand the type of issue the developer is correcting. Bigrams are a consecutive pair of two adjacent terms in a sentence. Furthermore, unlike unigrams (i.e., single terms), bigrams provide more context for the terms and thereby helps in reducing false presumptions.

\noindent\textbf{Approach:} First, we extract bigrams frequently occurring in SATD comments, following an approach similar to \citet{Peruma2021EMSEStackoverflow}. We utilize bigrams as they provide more context for the terms and thereby help in reducing false presumptions. Additionally, we also extract the refactoring-specific terminology in these comments. Our extraction process involves programmatically examining each source file associated with a refactoring commit and extracting the set of removed SATD comments. These comments are available in the diff hunks of the source dataset. We then normalize the text using a series of pre-processing activities, including converting the text to lowercase, expanding contractions, removing non-alphabetic characters, digits, ASCII punctuation characters, and stopwords (standard English terms and terms specific to the dataset). 

%To derive the set of frequently occurring bigrams, we examined each source file associated with a refactoring commit and extracted the set of removed SATD comments. These comments are available in the diff hunks of the source dataset (as mentioned in Section \ref{Section:experiment_design_source}). Before generating the bigrams, we normalized the text. The preprocessing steps consists of converting the text to lowercase, expanding contractions, removing non-alphabetic characters, digits, ASCII punctuation characters, and stopwords (standard English terms and terms specific to the dataset).

Table \ref{Table:rq2_bigram} shows the top ten frequently occurring bigrams in our dataset, while Table \ref{Table:rq3_sar} shows the dataset's top ten frequently occurring (stemmed) refactoring terms. Our analysis of these extracted bigrams and terms results in us proposing the below categories as rationales for technical debt repayment via refactoring.

\subsubsection*{\textbf{Error Handling}:}
Improving error handling is a frequent area of technical debt that developers address when refactoring their code. This is evident by the frequent occurrence of the bigrams: `catch block', `error handling', `exception handling', and `throw exception'. Looking at the removed comments, we observe text such as ``TODO: We could use a better strategy for error handling'' and ``TODO: Fix exception handling''. This shows that developers knowingly write code prone to errors or utilize generic (or auto-generated) error handling and then make necessary corrections when refactoring their code in later revisions of the codebase. For example, in Listing \ref{Listing:rq2_exception}, the developer removes code that always returns a null value.

\vspace{2mm}
\begin{minipage}{\linewidth}
\begin{lstlisting}[caption=Removal of erroneous code \cite{bigrams_exception01}., label=Listing:rq2_exception, firstnumber = last, escapeinside={(*@}{@*)}]
public Object nextElement()
{
    return null;// TODO: check exception handling
}
\end{lstlisting}
\end{minipage}

\subsubsection*{\textbf{Code \& Structural Improvements}:}
This category comprises of two sub-categories-- 1) Clean-up Activities and 2) Design Improvements. Clean-up activities can include removing temporary code or renaming identifiers, such as in the case the developer renames an identifier (e.g., ``TODO - Rename variable to commandId'' \cite{bigrams_cleanup}). From prior studies \cite{AlOmar2020ESWARefactoring,Peruma2019MobileSoftAndroid}, we know that clean-up is an activity associated with refactoring. Design-level changes include the removing and moving code (such as moving methods, method extraction, etc.) and data types changes. Further, the bigram `get rid' is mainly associated with code the developer needs to remove from the project. For instance, the comment ``TODO get rid of this cast''  is addressed by a Change Attribute Type operation \cite{bigrams_design}. From Table \ref{Table:rq3_sar}, the terms `mov', `remov', and `chang' are indicators of design-level operations.

\subsubsection*{\textbf{Feature Updates}:}
This category comprises of refactoring operations developers perform to incorporate feature changes in the system. For instance, the term`implement' is a placeholder for the developer to update the functionality, as in the example of ``FIXME: our implementation is flawed...'', in which the developer applies a Move Attribute refactoring in addition to a series of other code changes \cite{bigrams_func01}. In another example, we encounter the removal of the comment ``TODO: we can probably get a more efficient implementation...'' with an Extract Attribute refactoring \cite{bigrams_func02}. A similar placeholder term developers utilize is `add', like in the example ``TODO: Add method to extract...'', which is resolved by an Extract Method operation combined with other changes \cite{bigrams_func03}.

%While this data is interesting and further supports the fact that analyzing comments can help us understand the intention behind the refactoring, further study is required to create a more formal and exhaustive set of the causes of refactoring-based technical debt removal. 

\begin{table}
\fontsize{8}{9}\selectfont
\centering
\caption{Top ten frequent bigrams in SATD comments}
\vspace{-2.5mm}
\label{Table:rq2_bigram}
\begin{tabular}{@{}lrr@{}}
\toprule
\multicolumn{1}{c}{\textbf{Bigram}} & \multicolumn{1}{c}{\textbf{Count}} & \multicolumn{1}{c}{\textbf{Percentage}} \\ \midrule
catch block                         & 1,037                              & 6.03\%                                  \\
make thisdefault                    & 89                                 & 0.52\%                                  \\
make sure                           & 76                                 & 0.44\%                                  \\
get rid                             & 63                                 & 0.37\%                                  \\
error handling                      & 59                                 & 0.34\%                                  \\
pessimistic read                    & 56                                 & 0.33\%                                  \\
exception handling                  & 51                                 & 0.30\%                                  \\
throw exception                     & 50                                 & 0.29\%                                  \\
implement needed                    & 42                                 & 0.24\%                                  \\
add repository                      & 40                                 & 0.23\%                                  \\
\textit{Others}                     & 15,637                             & 90.91\%                                 \\ \bottomrule
\end{tabular}
\vspace{-3mm}
\end{table}

\begin{table}
	\fontsize{8}{9}\selectfont
\centering
\caption{Top ten frequent (stemmed) refactoring terms in SATD comments.}% associated with a single refactoring operation.}
\vspace{-2.5mm}
\label{Table:rq3_sar}
\begin{tabular}{@{}lrr@{}}
\toprule
\multicolumn{1}{c}{\textbf{Term}} & \multicolumn{1}{c}{\textbf{Count}} & \multicolumn{1}{c}{\textbf{Percentage}} \\ \midrule
add                        &  74                             & 20.32\%                                  \\
mov                    &   66                             & 18.13\%                                  \\
fix                          &    51                              & 14.01\%                                  \\
remov                            &   48                              & 13.18\%                                  \\
chang                   &    26                            & 7.14\%                                  \\
creat                 &   25                              & 6.86\%                                  \\
rewrit                 &    11                            & 3.02\%                                  \\
replac                     &    10                             & 2.74\%                                  \\
extend                   &    6                            & 1.64\%                                  \\
merg                      &    6                             & 1.64\%                                  \\
\textit{Others}                     &  41                           & \%                                 \\ \bottomrule
\end{tabular}
\vspace{-4mm}
\end{table}

% \begin{itemize}
%     \item Table \ref{Table:rq3_sar} depicts the top-10 SAR patterns in SATD comments that are associated with a single refactoring operation. 
%     \item Developers mostly utilized generic SAR terminology to indicate potential design problems.
%     \item Refactoring operation-related terms are rarely documented in technical debt comments and are among the lower ranked refactoring textual description. 
%     \item SAR-related categories (i.e., internal QA, external QA, and code smell) are not documented by developers in SATD comments.
% \end{itemize}

% \vspace{-1.5mm}
\begin{tcolorbox}[top=0.5pt,bottom=0.5pt,left=1pt,right=1pt]
\textbf{Summary for RQ2.} Analysis of SATD comments show that developers refactor to repay technical debt related to error handling, code optimization, and system features. 
\end{tcolorbox}

% \section{Related Work}
% \label{Section:related_work}

\section{Threats To Validity}
\label{Section:threats}
Though our dataset is limited to Java systems, the 76 systems are well established and widely utilized open-source systems. Furthermore, as we utilize an existing dataset, our analysis is limited to the data points contained within the source dataset. This includes the dataset authors' mechanisms/tools to identify SATD comments in the code  (i.e., presence of the terms ``TODO'', ``FIXME'', and ``XXX'') and mine refactoring operations. Furthermore, since the dataset does not provide a straightforward mechanism to map SATD comments to an identifier, our analysis of SATD is limited to the file and commit level and not at the identifier level. That said, the features, volume, and age of data are more extensive and more recent than similar datasets \cite{Lenarduzzi2019Dataset,Iammarino2019SATD}.

\section{Discussion \& Conclusion}
\label{Section:discussion}

As an exploratory study, our research aims to understand the extent of the relationship between technical debt repayment and code refactoring, and our preliminary findings show promise in further investigating these trends. Furthermore, we provide direction to research areas for supporting developers in designing and maintaining their code.

Our RQ1 findings confirm prior work showing that the removal of technical debt is frequently associated with refactoring actions \cite{Iammarino2019SATD,Perez2020TechDebt,Perez2020IST}. In contrast, we observe differences in the types of refactoring operations that frequently occur. That said, when comparing the results from prior studies, such as \cite{Iammarino2019SATD} and \cite{Zabardast2020SEAA}, we observe a slight consensus between the reported frequent operations. While it can be argued that the cause can be due to the datasets and mining tools, further research in this area is warranted, especially since we did not observe a significant difference between the operations applied to remove technical debt and general refactoring. 

Our RQ2 results provide interesting insight into the association of refactoring and technical debt repayment and support a further study to create a more formal and exhaustive set of causes. However, we see specific parallels with prior work. For instance, the category of repayment associated with features is also mentioned by \cite{Zampetti2018MSR}, while we share code \& structural improvements and error handling with \cite{Digkas2018SANER}. Additionally, our categories also share similarities with the refactoring motivation taxonomy proposed by \cite{AlOmar2020ESWARefactoring}.

Below, we discuss how the findings from our RQs support the community through a series of takeaways.

% \vspace{1.0mm}
\noindent\textbf{Takeaway 1: Support for robust error handling.}
Our findings of developers improving error handling in their code is an opportunity for tool and IDE vendors to provide automated support for detecting such shortcomings in the code. Furthermore, developers should also understand that auto-generated try-catch blocks must be customized as per the project specifications.% and not left as-is.

% \vspace{1.0mm}
\noindent\textbf{Takeaway 2: Improving the accuracy of refactoring recommendation tools/models.}
With the repayment of technical debt frequently occurring with refactoring activities, refactoring recommendation tools/models, such as identifier renaming \cite{Li2020ACM} and appraisal \cite{Peruma2021IDEAL}, can improve their accuracy by considering the occurrence of such SATD comments in their recommendation approach. Furthermore, there is a research opportunity for defining a standard set of composite refactorings to repay common types of debt since repayment usually involves applying complex changes \cite{Zampetti2018MSR}.

% \vspace{2mm}
%\subsection*{Future Work}
%As an exploratory study, we see promise in further investigating these trends in technical debt repayment through refactoring. Our future studies will expand our RQ 2 categorization through developer surveys to formalize and confirm a taxonomy of rationales for technical debt repayment via refactoring.

\bibliographystyle{ACM-Reference-Format}
\bibliography{references}

\end{document}